# Réflexions sur la puissance motrice du Soleil

## Reflections on the motive power of the Sun

Daniel S<small>UCHET</small>, Nathan R<small>OUBINOWITZ</small>, Jean François G<small>UILLEMOLES</small>

*Institut Photovoltaique d'Île de France, UMR-IPVF 9006, CNRS, École Polytechnique IPP, ENSCP PSL, Palaiseau, France.*

**R<small>ESUME</small>** *Source de chaleur par excellence, le Soleil rayonne vers la Terre une puissance dix mille fois supérieure aux besoins d'énergie de l'humanité. Tirer parti de cette manne énergétique nécessite cependant de parvenir à capter et à convertir la lumière. Cette conversion peut aujourd'hui être réalisée par plusieurs familles de technologies, à différents degrés de maturité : solaire photovoltaïque, thermique, à concentration… Si leurs applications sont différentes, toutes ces technologies doivent répondre à des contraintes fondamentales communes, et on peut comme Carnot « envisager dans toute sa généralité le principe de la production du mouvement par la chaleur » du Soleil. Mais contrairement aux « machines à feu », le couplage avec la source chaude est ici radiatif, ce qui ajoute des contraintes particulières dont il faut tenir compte. Dans cet exposé consacré aux machines radiatives, on retrouvera donc des termes familiers, mais aussi des expressions particulières qui donneront les clés de compréhension des technologies solaires.*

**A<small>BSTRACT</small>** *A quintessential source of heat, the Sun radiates toward the Earth a power ten thousand times greater than humanity's energy needs. Harnessing this energy bounty, however, requires capturing and converting sunlight. Today, this conversion can be achieved through several families of technologies at varying stages of maturity: photovoltaic solar, thermal, concentrated solar power, and more. While their applications differ, all these technologies must meet common fundamental constraints, and as Carnot proposed, one can 'consider in all its generality the principle of producing motion through heat' from the Sun. However, unlike traditional 'heat engines,' the coupling with the hot source here is radiative, introducing specific constraints that must be accounted for. In this presentation dedicated to radiative machines, you will encounter familiar terms as well as particular expressions that will provide the keys to understanding solar technologies*

**M<small>OTS-CLES</small>** Energie solaire, photovoltaïque, rayonnement, thermodynamique

**K<small>EYWORDS</small>** Solar energy, photovoltaics, radiation, thermodynamics

Dans l'introduction de ses *Réflexions*, Sadi Carnot évoque l'idée que la chaleur est à l'origine des « *grands mouvemens qui frappent nos regards sur la terre [...] : agitations de l'atmosphère, l'ascension des nuages, la chute des pluies* » etc. Si la suite de son œuvre se concentre sur des machines à feu alimentées par la houille et le charbon, la chaleur qui alimente ces phénomènes naturels ne provient pas de la combustion de ressources terrestres. Cette chaleur est en effet portée par rayonnement solaire qui atteint le vaisseau spatial Terre après avoir parcouru 150 millions de kilomètres. Pour compléter le projet de Carnot et envisager la conversion de la chaleur « *dans toute sa généralité* », on peut s'interroger sur la possibilité de transformer en puissance motrice la chaleur issue de ce feu Prométhéen.

Imaginons donc un absorbeur qui reçoit la chaleur du rayonnement solaire et fournit en échange du travail. La température de surface du Soleil est portée à environ 6 000 K par les réactions de fusion nucléaire entretenues par la gravité solaire. Pour une machine thermique opérant entre cette source chaude et un environnement froid à 300 K, le théorème de Carnot impose un rendement maximal de 95 %. Cette analyse ne tient cependant pas compte d'une spécificité du problème : l'échange d'énergie est ici radiatif, ce qui impose des contraintes supplémentaires pour la conversion de la chaleur reçue.

## 1. De la loi de Stefan-Boltzmann à la ressource solaire

Contrairement aux réservoirs de chaleur imaginés par Carnot, une source de rayonnement émet une quantité d'énergie qui dépend directement de sa température. Pour un émetteur idéal, c'est-à-dire un *corps noir* capable de rayonner dans toutes les longueurs d'onde, la puissance émise par une surface élémentaire dans une direction donnée s'exprime suivant la loi de Stefan-Boltzmann :

$$\phi_E(\Omega) = \frac{\cos\theta}{\pi} \times \sigma T^4$$

où $\sigma = 5.67 \times 10^{-8}$ W/m²/K⁴ est la constante de Stefan. La dépendance en $\cos\theta$ de la distribution angulaire caractérise la nature Lambertienne de l'émission. Du fait de ce facteur, la puissance surfacique émise dans un angle solide d'ouverture $\theta_0$ s'exprime comme $\phi_E = \sin^2\theta_0 \; \sigma T^4$. Pour une émission tous azimuths ($\theta_0 = \pi/2$) à la température de surface du Soleil (5 800 K), on peut évaluer à 65 MW/m² la puissance surfacique émise par le Soleil – historiquement, c'est par cette loi que Stefan réalise la première estimation de la température du Soleil en 1879.

Pour estimer la puissance totale reçue par un absorbeur, il faut additionner la puissance que chaque unité de surface de l'émetteur rayonne en direction de l'absorbeur. La somme de ces contributions peut être réécrite comme :

$$\phi_E = \sin^2\theta_s \times \sigma T^4$$

où $\theta_s$ est l'angle sous lequel la source est vue depuis l'absorbeur.

La quantité $S_{abs} \sin^2\theta_s$ est *l'étendue optique* du faisceau. Obtenue ici par des considérations géométriques, elle porte en réalité un sens thermodynamique puisqu'elle caractérise le volume du faisceau dans l'espace des phases (produit de sa surface dans l'espace réel et de son ouverture angulaire dans l'espace réciproque) et se trouve ainsi liée à l'entropie du rayonnement [1].

Depuis la Terre, le Soleil est vu sous un angle $\theta_s \simeq 16'$. La formule précédente permet d'estimer la *constante solaire*, puissance reçue par unité de surface au sommet de l'atmosphère : $P/S_{abs} \simeq 1360 \; W/m^2$. En tenant compte de l'absorption et de la réflexion de la lumière lors de traversée de l'atmosphère, cette puissance est réduite à 1 000 W/m² au niveau du sol, ce qui constitue la valeur de référence du rayonnement solaire utilisée pour caractériser l'efficacité des dispositifs de conversion.

Pour estimer la ressource énergétique, il faut également tenir compte de facteurs géométriques (alternance jour/nuit, saisons…) et météorologiques. La puissance moyennée sur l'année est alors de l'ordre de 150 W/m2. Sur les terres émergées (30 % de la surface terrestre), cette puissance représente environ 20 000 TW, plus de 1 000 fois la consommation primaire de l'humanité [2].

## 2. De la loi de Kirchhoff à la machine de Müser

Pour tirer parti de cette manne énergétique, il faut que l'absorbeur absorbe effectivement le rayonnement incident. Or la capacité d'un corps à absorber la lumière est directement liée à sa capacité à émettre un rayonnement. La relation entre ces deux propriétés est établie par la loi de Kirchhoff. Cette loi est très simple à exprimer, mais nécessite de bien définir les quantités considérées.

D'une part, la capacité d'absorption est décrite par l'absorptivité A. Si le corps était éclairé par un faisceau monochromatique de longueur d'onde λ, alors il absorberait une fraction A(λ) de la puissance incidente (le reste étant réfléchi ou transmis au travers du système).

D'autre part, la capacité d'émission est décrite par l'émissivité E. Si le corps est à l'équilibre thermique à une température T, alors il émet un rayonnement a priori dans tout le spectre. À la longueur d'onde λ, la puissance rayonnée par le corps est une fraction E(λ) de celle qu'émettrait un corps noir porté à la même température.

Un corps noir absorbe tous les rayonnements qui l'atteignent (A(λ) = 1) et émet lui-même un rayonnement de corps noir (E(λ) = 1). La loi de Kirchhoff généralise cette identité entre absorptivité et émissivité :

$$A(\lambda) = E(\lambda)$$

Elle ne dit pas que le corps émet à chaque longueur d'onde autant que ce qu'il absorbe – elle relie la capacité à émettre (comparé à un corps noir) à la capacité à absorber (si un rayonnement était incident).

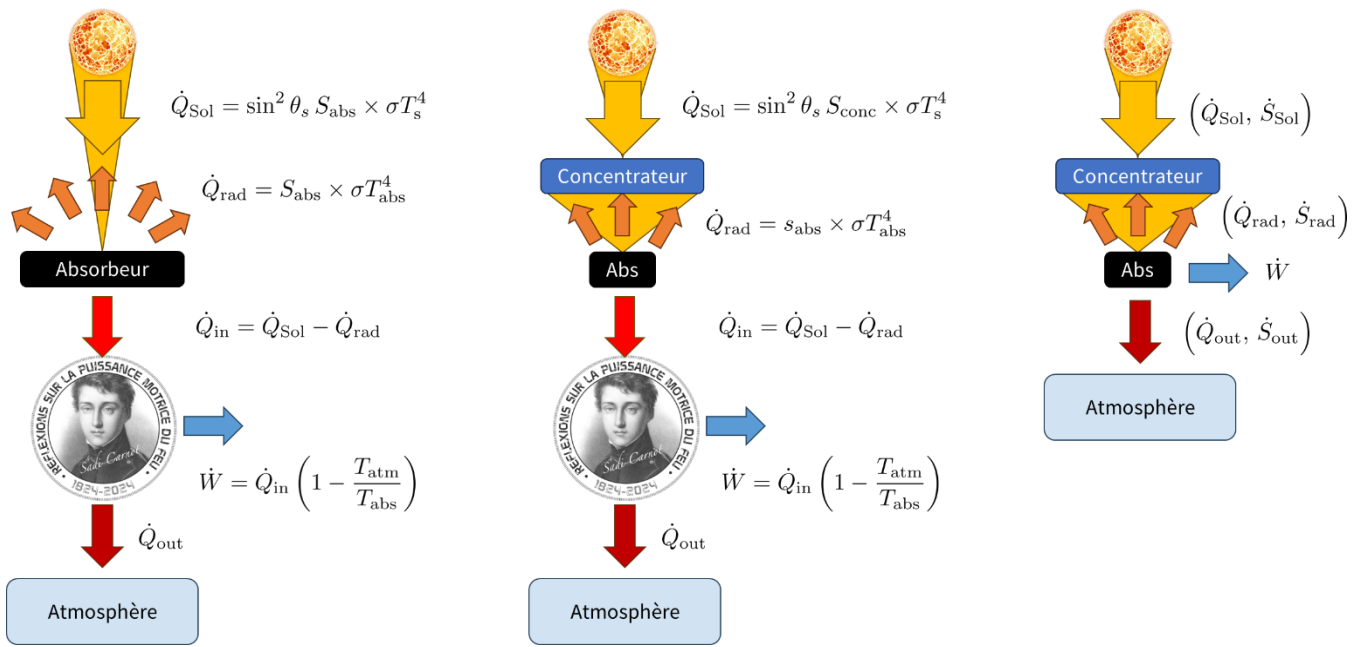

**Figure 1 —** bilan de la machine de Müser avec et sans concentration optique (gauche, milieu) et de la machine de Landsberg (droite).

Armé des lois de Stefan et de Kirchhoff, on peut estimer le rendement de conversion d'une première machine solaire (machine de Müser), composée d'un absorbeur parfait (corps noir) qui fournit de la chaleur à un moteur idéal (moteur de Carnot). Le bilan énergétique de ce système est représenté en figure 1 (gauche) et son rendement s'écrit :

$$\eta_{Muser} = \frac{\dot{W}}{\dot{Q}_{sol}} = \left(1 - \frac{1}{\sin^2 \theta_s} \frac{T_{abs}^4}{T_{sol}^4}\right)\left(1 - \frac{T_{amb}}{T_{abs}}\right)$$

Ce rendement est limité à moins de 7 % par un compromis désastreux (figure 2 gauche). Pour que le moteur fonctionne efficacement, la chaleur doit être fournie à une température élevée ; mais si l'absorbeur est chaud, il émet beaucoup de rayonnement, et seule une petite fraction de la puissance absorbée reste disponible pour être transmise au moteur.

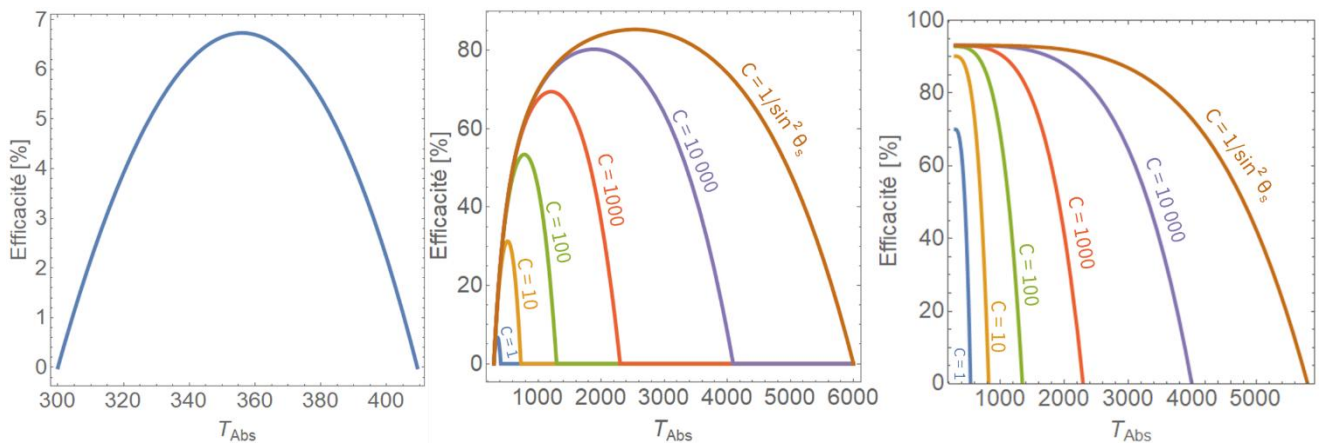

**Figure 2 —** rendement de la machine de Müser avec et sans concentration optique (gauche, milieu) et de la machine de Landsberg (droite).

Une première stratégie pour dépasser ce compromis consiste à découpler absorption et émission en séparant la surface qui collecte de la lumière et celle qui l'absorbe effectivement. C'est la voie du solaire à concentration.

Considérons à présent que la lumière est reçue sur une lentille ou un miroir convergent, de surface $S_{coll}$, qui redirige la lumière vers un absorbeur (figure 2 centre). La surface $S_{abs}$ de cet absorbeur doit couvrir l'image du Soleil formée par le système optique ; elle est plus petite que $S_{coll}$ d'un facteur $C = S_{coll}/S_{abs}$, appelé *facteur de concentration*.

En reprenant le bilan précédent, on écrit à présent le rendement de la machine comme :

$$\eta_{Muser} = \frac{\dot{W}}{\dot{Q}_{sol}} = \left(1 - \frac{1}{C \times \sin^2 \theta_s} \frac{T_{abs}^4}{T_{sol}^4}\right)\left(1 - \frac{T_{atm}}{T_{abs}}\right)$$

Cette efficacité est d'autant plus grande que le facteur de concentration est élevé (figure 2 centre). Cependant, le facteur de concentration ne peut être arbitrairement grand. Contrairement à ce que laisse penser une approche trop simple d'optique géométrique, la taille minimale de l'image réalisée par le système optique dépend de la taille du système de collecte. Ce résultat vient de la conservation de l'étendue optique du faisceau :

$$S_{coll} \sin^2 \theta_s = S_{abs} \sin^2 \theta_{coll}$$

où $\theta_{coll}$ est l'angle sous lequel le dispositif de collecte est vu depuis l'absorbeur. Ainsi, l'image la plus petite est obtenue lorsque le système de collecte couvre l'intégralité du ciel vu depuis l'absorbeur ; on a alors $C = 1/\sin^2 \theta_s \simeq 46\,000$. À pleine concentration, le rendement peut atteindre 86 % – à condition de trouver un absorbeur qui supporte des températures de l'ordre de 3 000 °C.

On retrouve ici le sens thermodynamique de l'étendue optique : s'il était possible d'atteindre des concentrations plus élevées, l'absorbeur pourrait atteindre des températures supérieures à celle du Soleil, en contradiction avec le second principe. (Pour la même raison [3], on ne pourra jamais concentrer la lumière de la lune suffisamment pour allumer un feu avec une loupe géante en pleine nuit !).

Le solaire à concentration a été utilisé dès les toutes premières tentatives pour obtenir du travail à partir du rayonnement solaire. Ainsi, le moteur solaire d'Augustin Mouchot, présenté à Napoléon dans les années 1860, alimente une machine à vapeur par un grand miroir parabolique – on comprend à présent que la concentration n'est pas simplement une amélioration, mais une condition sine qua non du fonctionnement de la machine.

À l'heure actuelle, la plupart des projets utilisent un champ de miroirs plans pour renvoyer la lumière vers le sommet d'une tour centrale où circule un liquide caloporteur. Avec des facteurs de concentration de l'ordre de 1 000, le système atteint des températures supérieures à 800 °C, largement suffisantes pour alimenter une turbine et produire une puissance électrique. Ces installations nécessitent cependant de grandes surfaces ;

elles doivent suivre la course du Soleil dans le ciel et ne sont adaptées qu'aux régions avec de faibles couvertures nuageuses – la lumière diffuse ne pouvant être efficacement concentrée.

## 3. Du théorème de Carnot au rendement de Landsberg

Il est cependant possible d'obtenir du travail directement à partir du rayonnement solaire, sans passer par l'astuce de la concentration. Dans l'approche de Müser, on a considéré que l'énergie $\dot{Q}_{in}$ cédée par l'absorbeur était fournie au moteur intégralement sous forme de chaleur. Mais une partie de cette énergie peut en réalité être donnée directement sous forme de travail.

Pour s'en convaincre, il faut suivre les mêmes étapes que pour établir le rendement de Carnot et compléter le bilan d'énergie d'un bilan d'entropie (figure 1 droite). Le flux surfacique d'entropie émis par un corps noir s'exprime comme :

$$\phi_S(\Omega) = \frac{\cos\theta}{\pi} \times \frac{4}{3}\sigma T^3$$

Ainsi, l'absorbeur reçoit et émet un flux d'énergie et un flux d'entropie. Il doit céder une quantité de chaleur $\dot{Q}_{out} = T_{atm}(\dot{S}_{sol} - \dot{S}_{rad})$ pour évacuer l'entropie excédentaire ; l'énergie qui n'a pas été perdue par radiation ou par chaleur peut être utilisée comme travail. On trouve ainsi le rendement de Landsberg :

$$\eta_{Lansberg} = \frac{\dot{Q}_{sol} - \dot{Q}_{rad} - \dot{Q}_{out}}{\dot{Q}_{sol}} = 1 - \frac{4}{3}\frac{T_{atm}}{T_{sol}} + \frac{1}{C \times \sin^2\theta_{sol}}\frac{T_{atm}^4}{T_{sol}^4}$$

qui atteint 70 % sans concentration[1] et 93,1 % à pleine concentration (figure 2 droite). Cette valeur ultime de la conversion de l'énergie solaire à la limite réversible est l'analogue du théorème de Carnot pour les échanges radiatifs.

## 4. De la loi de Planck aux cellules photovoltaïques

Jusqu'à présent, tout comme Carnot, nous n'avons caractérisé la source d'énergie que par sa température. Outre son intensité, l'énergie lumineuse est caractérisée par son spectre, c'est-à-dire l'intensité à chaque longueur d'onde. Cette spécificité offre une autre voie pour réaliser la conversion de l'énergie solaire. Pour l'explorer, il nous faut compléter la loi de Stefan par la loi de Planck, qui décrit la répartition du rayonnement dans le spectre électromagnétique en fonction de la température de l'émetteur.

---

[1] Pour mener le calcul à faible concentration, il faut également tenir compte des échanges radiatifs avec le reste du ciel à 300 K.

À la fin du XIXe siècle[2], deux modèles parviennent à capturer les comportements asymptotiques du rayonnement du corps noir. Fort de sa connaissance des ondes acoustiques, Rayleigh s'intéresse à la densité de modes de l'onde lumineuse. En considérant ces modes comme tous également peuplés il trouve une expression qui décrit correctement la forme du rayonnement à basse énergie, mais diverge à haute énergie (c'est la « catastrophe ultraviolette », comme l'appellera bien plus tard Paul Ehrenfest). De son côté, Wien part de la thermodynamique classique et des observations expérimentales pour proposer une loi d'échelle qui décrit remarquablement le spectre à haute énergie, mais s'avère erronées lorsque de nouvelles mesures améliorent la résolution à basse énergie.

Max Planck propose alors une amélioration phénoménologique de la loi de Wien [4]. Il se représente le rayonnement thermique à l'équilibre avec des oscillateurs mécaniques et s'intéresse à la distribution d'entropie des oscillateurs – ou plus exactement, à la dérivée seconde de l'entropie par rapport à l'énergie « car cette quantité a un sens physique plus évident » (sic). L'intuition de Planck lui suggère la relation :

$$\left(\frac{\partial^2 S}{\partial U^2}\right)^{-1} = \alpha U + \beta U^2$$

C'est en intégrant cette relation qu'il écrit pour la première fois la loi qui porte aujourd'hui son nom. Ainsi, l'établissement de la loi de Planck, point de départ de la mécanique quantique, passe historiquement par l'entropie, formalisation par Clausius des idées introduites par Sadi Carnot dans ses Réflexions !

Dans une réécriture plus moderne, on écrit la loi de Planck comme :

$$\phi_E(h\nu, \Omega) = \frac{\cos\theta}{4\pi^3 \hbar^3 c^2} \frac{(h\nu)^3}{e^{\frac{h\nu}{k_B T}} - 1}$$

où $h = 6\,10^{-34} J.s$ est la constante de Planck, $\hbar = h/2\pi$ la constante de Planck réduite et $k_B = 1.38\,10^{-23}\,J/K$ est la constante de Boltzmann. En intégrant sur toutes les longueurs d'onde, on retrouve bien la loi de Stefan citée plus haut.

Cette distribution offre une nouvelle voie pour contourner le compromis imposé par la loi de Kirchhoff. Le spectre solaire est maximal vers 500 nm, tandis que le rayonnement émis à température ambiante est situé autour de 10 µm. Ainsi, on peut imaginer un absorbeur capable de capter et d'émettre la lumière à courte longueur d'onde (où les photons solaires sont nombreux et l'émission faible), mais transparent aux grandeurs longueurs d'onde (où l'émission serait significative et l'éclairement solaire faible).

---

[2] Pour plus de détails sur les aspects historiques, voir l'excellent site de Gilles Montambaux https://gilles.montambaux.com/histoire-physique/.

Depuis la découverte expérimentale de l'effet photovoltaïque par Edmond Becquerel [5], ce filtrage spectral est réalisé en utilisant comme absorbeur un semi-conducteur, dont l'absorptivité ressemble idéalement à un créneau $A(\lambda) = \theta(hc/\lambda - E_g)$. Ainsi, seuls les photons plus énergétiques que le gap seront absorbés et potentiellement émis.

Dans un premier modèle simple [6], on peut considérer que chaque photon d'énergie $h\nu \geq E_g$ promeut un électron de la bande de valence vers la bande de conduction ; que chaque électron ainsi excité est extrait de l'absorbeur au-dessus du gap, puis réinjecté en dessous du gap. On suppose ainsi que les 4 fonctions fondamentales du photovoltaïques (absorption optique, durée de vie des excitations, transport des électrons et sélectivité de l'extraction) sont parfaitement réalisées [7]. La puissance délivrée par ce dispositif s'écrit alors comme :

$$P(E_g) = S_{abs} \sin^2 \theta_{sun} \int_{h\nu=E_g}^{+\infty} \frac{\phi_E^{sol}(h\nu)}{h\nu} E_g \, d(h\nu)$$

Cette puissance est nulle si le gap est trop grand (car trop peu de photons sont absorbés et le courant est nul) ou si le gap est trop petit (car l'énergie récupérée par électron extrait est nulle) – il existe donc un gap optimal pour convertir le rayonnement solaire.

Ce modèle peut être rendu rigoureux en tenant compte des recombinaisons radiatives rendues inévitables par la loi de Kirchhoff (tous les électrons excités ne seront donc pas extraits), et en distinguant dans l'énergie totale obtenue en extrayant un électron la part correspond à un travail (i. e. la tension électrique aux bornes du dispositif). Cette analyse plus détaillée met en évidence que l'origine du travail produit par une cellule solaire n'est pas un déséquilibre de température, mais un déséquilibre des populations électroniques, qui se traduit également dans le rayonnement émis (*loi de Planck généralisée*) [8]. On trouve alors que le rendement maximal, obtenu avec un gap de 1,1 eV, atteint environ 30 % [9] (figure 3).

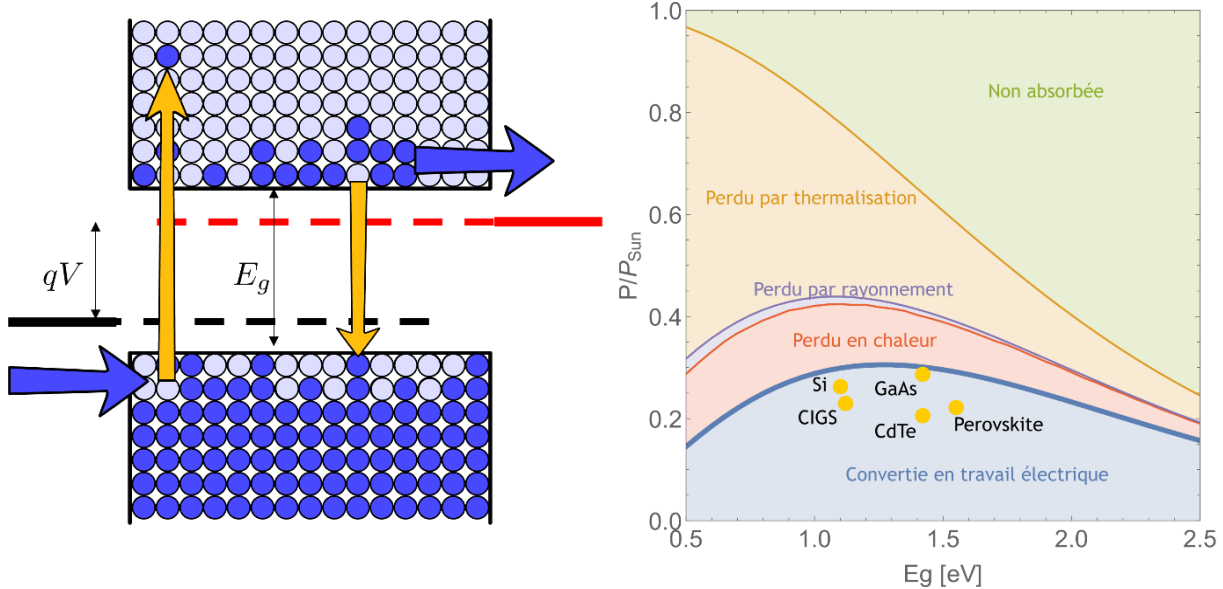

**Figure 3 —** (gauche) lors de l'absorption de photons d'énergie supérieure au gap Eg, des électrons de la bande de valence sont excités vers la bande de conduction. Ces « porteurs » sont ensuite acheminés vers le bord du dispositif pour les faire travailler dans un circuit électrique. Lors de ce déplacement, plusieurs pertes ont lieu : relaxation (ou « thermalisation ») au sein de la bande de conduction, recombinaison vers la bande de valence (rayonnement), perte sous forme de chaleur lors du transfert au circuit électrique. (droite) répartition des pertes dans une cellule photovoltaïque en fonction du gap. Les points jaunes indiquent les performances record obtenues en laboratoire pour différents matériaux absorbeurs.

Ces estimations simples ouvrent trois perspectives qui constituent autant de voies de recherche [10]. On peut chercher à caractériser et à fabriquer des objets capables de s'approcher de cette limite, avec des contraintes de conception et de fabrication de plus en plus fortes[3] (coût économique, éco-conception, propriétés mécaniques ou esthétiques…). On peut entreprendre de dépasser cette limite pour se rapprocher du rendement de Landsberg, en jouant sur l'architecture des dispositifs (multijonctions), la conversion des photons incidents et les propriétés électroniques de l'absorbeur. On peut enfin généraliser la question à d'autres sources de rayonnement que le Soleil et d'autres sources froides que l'air ambiant [11].

Dans tous les cas, l'approche thermodynamique reste pertinente, et vient à être complétée par des concepts issus de l'optique, de l'électronique et de la science des matériaux.

---

[3]Pour plus de détails sur le solaire photovoltaïque sur le terrain et en particulier en France, voir le guide académique https://solairepv.fr/.